\documentclass[prl,twocolumn,nofootinbib,nobibnotes,preprintnumbers,floatfix]{revtex4-1}
\usepackage{amssymb,amsmath,graphicx}
\usepackage[dvipsnames]{xcolor}
\usepackage[colorlinks,allcolors=NavyBlue]{hyperref}
\usepackage{microtype}

\newcommand{\be}{\begin{eqnarray}}
\newcommand{\ee}{\end{eqnarray}}

\newcommand{\benum}{\begin{enumerate}}
\newcommand{\eenum}{\end{enumerate}}
\newcommand{\bi}{\begin{itemize}}
\newcommand{\ei}{\end{itemize}}

\newcommand{\Eq}[1]{Eq.~(\ref{#1})}  
\newcommand{\brac}[2]{ \left( \frac{#1}{#2} \right) }

\begin{document}

\preprint{FERMILAB-PUB-23-349-T}

% \title{ Detecting Dark Matter Decays with the Line Emission Mapper}
\title{Probing The Longest Dark Matter Lifetimes with the Line Emission Mapper}

\author{Gordan Krnjaic$^{1,2,3}$}
\thanks{krnjaicg@fnal.gov, http://orcid.org/0000-0001-7420-9577}

\author{Elena Pinetti$^{1,2}$}
\thanks{epinetti@fnal.gov, http://orcid.org/0000-0001-7070-0094}

\affiliation{$^1$Fermi National Accelerator Laboratory, Theoretical Astrophysics Group, Batavia, Illinois 60510, USA}
\affiliation{$^2$University of Chicago, Kavli Institute for Cosmological Physics, Chicago, Illinois 60637, USA}
\affiliation{$^3$University of Chicago, Department of Astronomy and Astrophysics, Chicago, Illinois 60637, USA}

\date{\today}

\begin{abstract}
In the next decade, the proposed Line Emission Mapper (LEM) telescope concept is poised to 
revolutionize Galactic and extragalactic X-ray sensitivity. 
The instruments aboard LEM feature unprecedented eV scale energy resolution and an effective area of 1600 cm$^2$ at 0.5 keV. Such features are ideally suited to 
explore decaying dark matter candidates that predict X-ray signals, including axion-like particles
and sterile neutrinos. 
We present the first forecast of LEM sensitivity to dark matter decays and find sensitivity  
to lifetimes beyond $\sim 10^{32}$ s in the keV range, surpassing current limits by
several orders of magnitude. Notably, our results show that LEM will be the first ever instrument
to probe such long dark matter lifetimes  in any mass range for any decay channel.

%  decays
% line search aimed at the galactic 
% can probe dark matter decays to photon lines with unprecedented sensitivity, 
% exceeding current limits by several orders of magnitude in the few 100 eV - few keV 
% mass range. 

\end{abstract}

\maketitle

\section{Introduction}

While the evidence for the existence of dark matter (DM) is overwhelming, 
its particle nature remains elusive (see \cite{Bertone:2016nfn} for a review). 
If DM particles are metastable on cosmological timescales,
their visible decay products may be observed as excesses over 
astrophysical backgrounds with terrestrial and space-based telescopes \cite{Hooper:2018kfv,Safdi:2022xkm}. This technique is particularly promising 
for keV  scale DM particles whose decays yield X-ray lines, including well motivated
models of axion-like particles \cite{Adams:2022pbo} and sterile neutrinos \cite{Abazajian:2017tcc}.

The Line Emission Mapper (LEM) is a proposed next-generation X-ray telescope consisting of two instruments aboard: an imaging calorimeter and a grazing-incidence mirror with a large effective area (1600 cm$^2$ at 0.5 keV)  \cite{Kraft:2022mnh}. Both instruments are optimized for detecting soft X-ray emission in the 0.2-2 keV energy interval. This innovative concept includes a large angular field of view (30' $\times$ 30') with a 10''-15'' angular resolution. LEM will have a remarkable grasp of $140\times 10^4$ cm$^2$ arcmin$^2$ at 0.5 keV, nearly three orders of magnitude better than the future X-ray telescope XRISM Resolve, which is planned to be launched in late 2030s \cite{team:2023gzc}. 

One of the most striking and distinctive features of this mission design is the exceptional spectral resolution of 1-2 eV, which will allow numerous emission lines to be detected with unparalleled precision. From a DM perspective, this energy resolution will have the potential to identify a photon line of DM origin as well as better characterize the astrophysical background of our particle signal. This mission plan will be submitted during the NASA 2023 Astrophysics Probes call for proposals, with the goal of being launched in 2032 \cite{Kraft:2022mnh}. This timeline makes it even more important to forecast the sensitivity of this next-generation telescope to previously unexplored DM parameter space.

To estimate the order of magnitude of LEM sensitivity to the DM lifetime, consider the approximation
in which DM particles of mass $m$ decay to nearly monochromatic photons with lifetime $\tau$ in the Galaxy. For an instrument with $N_{\rm bg}$ expected background events in the energy bin containing $E_\gamma \approx m/2$,
the DM lifetime $\tau$ for which the signal exceeds the 2$\sigma$ statistical uncertainty satisfies
\be
\tau \sim 10^{31} {\rm s} \brac{\rm keV}{m} \! \brac{t_{\rm obs}  }{10 \rm \, Ms}
\!\!\brac{A_{\rm eff}}{10^3 \, \rm cm^2}  \! \brac{\Delta \Omega}{10^{-4} \rm sr} \!\!
\sqrt{ 
\frac{10^4}{N_{\rm bg}}} , 
\nonumber
\ee
where $A_{\rm eff}$ is the effective area, $t_{\rm obs}$ is the duration of the observation, and we have used \Eq{eq:fluxgal} with $D \approx  28$, rescaled
by the field of view $\Delta \Omega$ of the telescope (see below).
Note that for these LEM inspired values, this lifetime reach exceeds the longest DM 
lifetime limits ever achieved in any channel at any DM mass; the current record is $\sim 10^{30}$ s
from observations of ultra-high-energy cosmic-rays \cite{Das:2023wtk}. %previous X-ray line searches \cite{AxionLimits,Blanco:2018esa}. 

In this {\it Letter}, we explore the LEM sensitivity to DM decays that yield observable photon lines.  
Our analysis covers the Galactic flux from DM decays in our halo and the 
extragalactic emission accumulated over all cosmic redshifts.

\section{Dark Matter Decay Formalism}

\begin{figure}
    \centering \hspace{-0.45cm}
    \includegraphics[width= \linewidth]{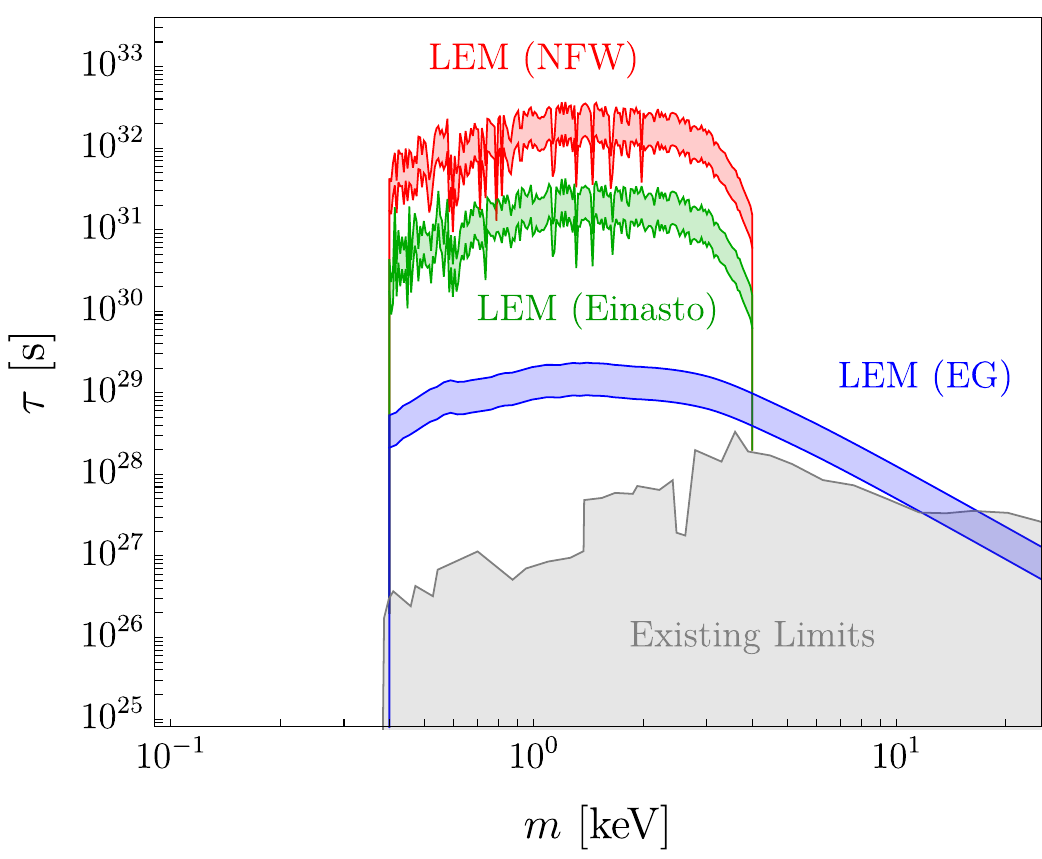}~~
    \caption{  Projection for LEM sensitivity to DM diphoton decays in the lifetime vs. mass plane. Here we use the effective area from Ref.  \cite{Kraft:2022mnh}, assuming an exposure time of 20 Ms.
    The red band shows the LEM 2-5 $\sigma$ sensitivity range assuming only decays from the 
    Galaxy with an NFW profile. The green band shows the same result with an Einasto profile, and the blue band shows the 
    corresponding projections for the 
    extragalactic contribution.  Also shown are existing 
    limits on the DM mass and lifetime assuming diphoton decays 
    \cite{Cadamuro:2011fd,AxionLimits}.
    }
    \label{fig:lifetime}
\end{figure}

\begin{figure*}
    \centering \hspace{-0.6cm}
        \includegraphics[width= 0.48\linewidth]{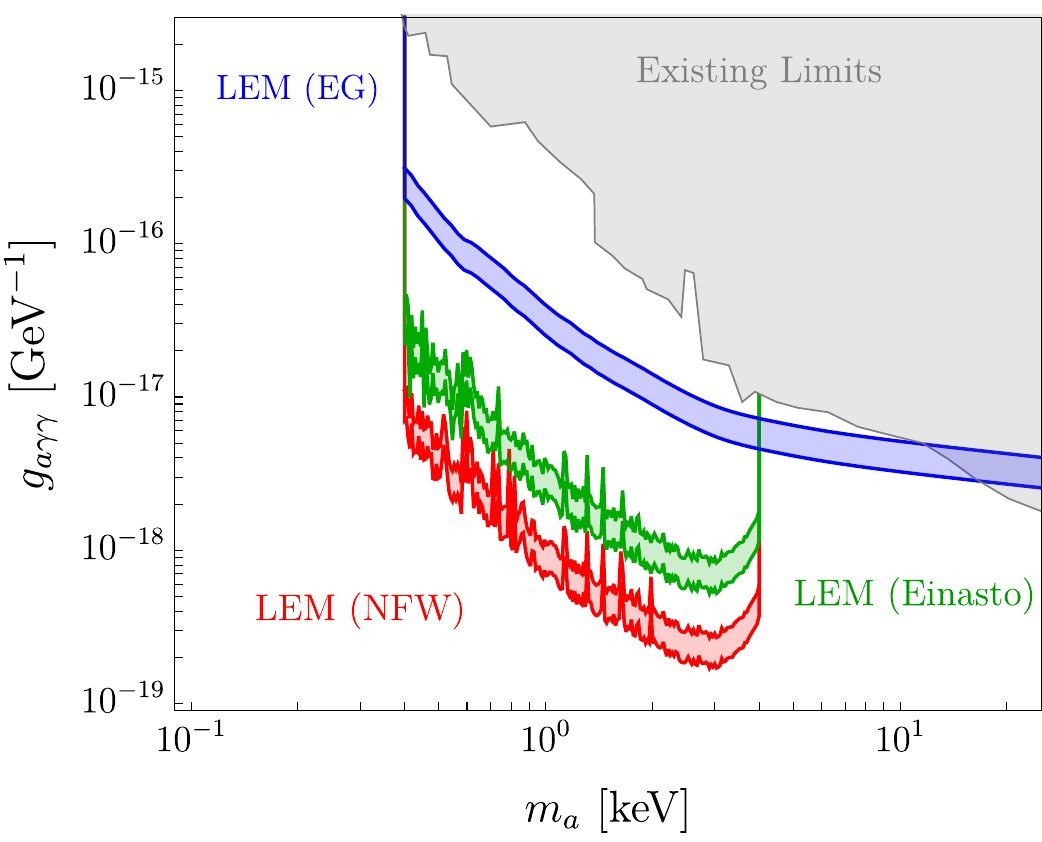}~~~
        \includegraphics[width= 0.48\linewidth]{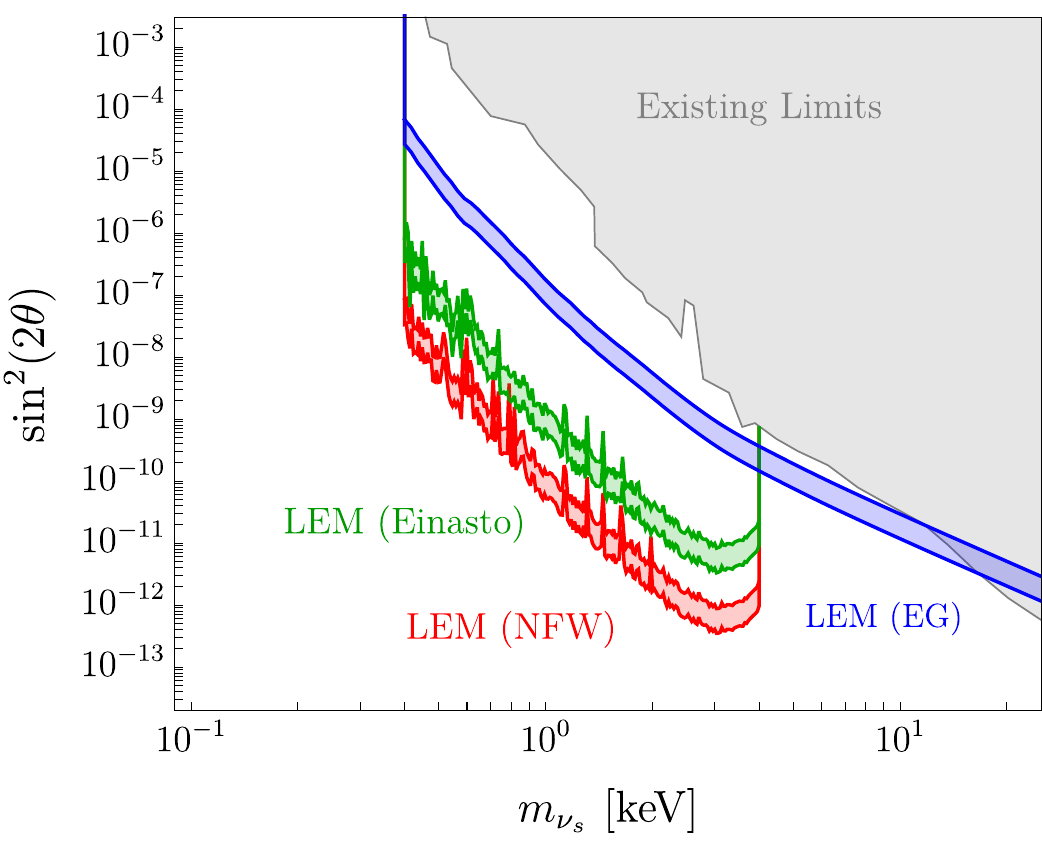}
    \caption{ {\bf Left: } Same analysis as in  Fig. \ref{fig:lifetime}, but interpreted as a limit on the diphoton coupling $g_{a\gamma\gamma}$ of an axion-like particle $a$ of mass $m_a$ with lifetime 
    given in \Eq{eq:tau_alp} and existing limits taken from Refs. 
    \cite{Cadamuro:2011fd,AxionLimits}.
     {\bf Right:} Same constraints as the left panel, but interpreted as limits on sterile neutrino
     dark matter with mixing angle $\theta$ and mass $m_{\nu_s}$ with lifetime from \Eq{eq:tau_nus}.
    }
    \label{fig:models}
\end{figure*}

\subsection{Halo Decays: DM Rest Frame}

In natural units with $\hbar = c = 1$, the terrestrial photon flux from Galactic DM decays can be written \cite{Hooper:2018kfv}
\be
\label{eq:fullflux}
\frac{d\phi}{dE_\gamma} = \frac{1}{4\pi m \tau} \frac{dN_\gamma}{dE_\gamma} \int_{\Delta \Omega} d \Omega \int_{\rm los} ds \, \rho(r),
\ee
where $m$ is the DM mass, $\tau$ is its lifetime, $dN_\gamma/dE_\gamma$ is the differential photon flux per decay
event, $d\Omega = d\ell  db \cos b$ is the solid angle 
in terms of Galactic latitude $b$ and longitude $\ell$, 
$\Delta \Omega$ is the instrumental field of view, $\rho$ is the DM density profile, and 
\be
r^2 = r_\odot^2 + s^2 - 2  r_\odot s \cos b \cos \ell
\ee
is the distance to the Galactic center (GC), $r_\odot = 8.25$ kpc is the solar distance from the GC, and $s$ is the line of sight coordinate. We can also
compactly write \Eq{eq:fullflux}
\be
\label{eq:fluxgal}
\frac{d\phi}{dE_\gamma} =  \frac{r_\odot \rho_\odot}{4\pi m\tau} \frac{dN_\gamma}{dE_\gamma} D,
\ee
where the dimensionless $D$ factor is 
\be
D \equiv \int_{\Delta \Omega} d\Omega \int_{\rm los} \frac{ ds }{r_\odot} \frac{ \rho(r) }{\rho_\odot}~,
\ee
and we have integrated over the field of view. 

To calculate the $D$ factor, we model the DM density using a Navarro-Frenk-White (NFW) profile \cite{Navarro:1996gj}
\be
\rho (r) = \dfrac{  \rho_0 }{ (r/r_s) \left( 1 + r/r_s \right)^2 },
\ee
where $r_s = $ 11 kpc is the scale radius \cite{Karukes_2019}.\footnote{Note that to improve the numerical stability of our $D$-factor integration, we remove the innermost region of the
Galactic center between $r= 0$ and  
$r = 10^{-4}$ kpc, where the NFW profile is sharply peaked, so our results should be regarded as conservative. }
To assess the sensitivity of our results to the choice of the halo profile, we also consider the Einasto profile \cite{1965TrAlm...5...87E}
\be
\rho(r) = \rho_0 \exp \left[ -(r/r_s)^a  \right],
\ee
where $a = 0.91$ and $r_s = 3.86$ kpc is the scale radius \cite{ou2023dark}.
For both profiles, the normalization parameter $\rho_0$ is chosen to recover the local DM density $\rho_\odot = $ 0.43 GeV cm$^{-3}$ \cite{Karukes_2019}.

Our signal of interest consists of a narrow X-ray line arising from a two-body DM decay 
process $\chi \to \gamma f$, where $f$ is any particle. 
In both cases, the final state photons emerge nearly monochromatically with $E_\gamma = m/2$
in the rest frame of the decaying particle
\be
\label{eq:delta_rest}
\brac{dN_\gamma}{dE_\gamma}_{\rm rest} = \delta\left( E_\gamma - \frac{m}{2}\right),
\ee
where $\delta$ is the Dirac delta function and we append an overall factor of 2 for the case where $f = \gamma$ and two photons are produced per decay.

\subsection{Halo Decays: Observer Frame}
As noted in Ref. \cite{Dessert:2023vyl}, for instruments with fine grained energy resolution, the DM line signal is affected by Doppler broadening due to two effects: the DM velocity dispersion and the line-of-sight solar velocity with respect to the Galactic reference frame.
Since LEM is projected to have unprecedented $\sim 2$ eV energy resolution, the energy deposited 
in the detector will be spread across several bins and the nature of the smearing depends on both
the Galactic coordinates of the decay event and the velocity of the observer.

To account for the effect of  DM velocity smearing, 
we first perform a thermal average in \Eq{eq:fullflux} 
taking the observer to be at rest with respect to the Galaxy
\be
\label{eq:therm}
\left(\frac{d\phi}{dE_\gamma}\right)_{\rm \! gal} \!\!= \frac{1}{4\pi m \tau} \int_{\Delta \Omega} \! d\Omega
\! \int_{\rm los} \! ds \, \rho(r) \! \int du \, f_{\rm gal}(u) \,\frac{dN_\gamma}{dE_\gamma}, \,
\ee
where $u \equiv \vec v\cdot \hat n$ is the DM velocity $\vec v$ projected onto the 
line-of-sight unit vector 
\be
\label{eq:nhat}
\hat{n} = (\cos b \cos \ell, \cos b \sin \ell, \sin b)~,
\ee
 so the energy distribution is now Doppler shifted in the 
 galaxy frame
\be
\label{eq:delta}
\frac{dN_\gamma}{dE_\gamma} = \delta\left( E_\gamma - \frac{m}{2}(1+u)\right).
\ee
In the Galactic rest frame,  the line-of-sight velocity distribution  can be written as
\begin{equation}
f_{\rm gal}(u) = \dfrac{1}{\sqrt{\pi}  v_0} e^{-u^2/v_0^2}~,
\end{equation}
where we assume a constant velocity dispersion $v_0 \approx 220$ km/s.\footnote{
 As found in Ref. \cite{Dessert:2023vyl}, line broadening effects in
the keV range are insensitive to the spatial dependence of $v_0$, so we adopt the  
 constant local value  throughout.}
Performing
the $u$ integration from \Eq{eq:therm} in the galaxy frame 
with the delta function from \Eq{eq:delta} yields to the thermally
averaged energy distribution 
\be
g_{\rm gal}(E_\gamma) 
&\equiv&
\int du \, f_{\rm gal}(u) \frac{dN_\gamma}{dE_\gamma} \nonumber \\
&=&
\dfrac{2}{\sqrt{\pi}  mv_0} \exp \! \left[{-\dfrac{4}{m^2 v^2_0 } \!\left( E_\gamma  -  \dfrac{m}{2} \right)^2}\right], 
\ee
where the observer velocity is $\vec v_\odot = (0, v_\odot, 0)$ with $v_\odot$ = 220 km/s,
and there is implicit $(b, \ell)$ dependence from $\hat n$ in \Eq{eq:nhat}.

Finally, to account for the additional Doppler shift from the observer's velocity,
the photon energy distribution becomes
\be
g_{\rm obs}(E_\gamma)  = g_{\rm gal}[E_\gamma(1+\vec v_\odot \cdot \hat n)],
\ee
and the flux in the observer frame is now 
\be
\label{eq:therm_obs}
\left(\frac{d\phi}{dE_\gamma}\right)_{\rm \! obs} \!\!= \frac{1}{4\pi m \tau} \int_{\Delta \Omega}  d\Omega
 \int_{\rm los}  ds \, \rho(r) \, g_{\rm obs}(E_\gamma)~,
\ee
which we integrate over our energy range of interest to compute the 
DM decay signal. 

\subsection{Extragalactic DM Decay Flux}

\begin{figure*}
    \centering
    \hspace{-0.6 cm}
    \includegraphics[width= 0.45\linewidth]{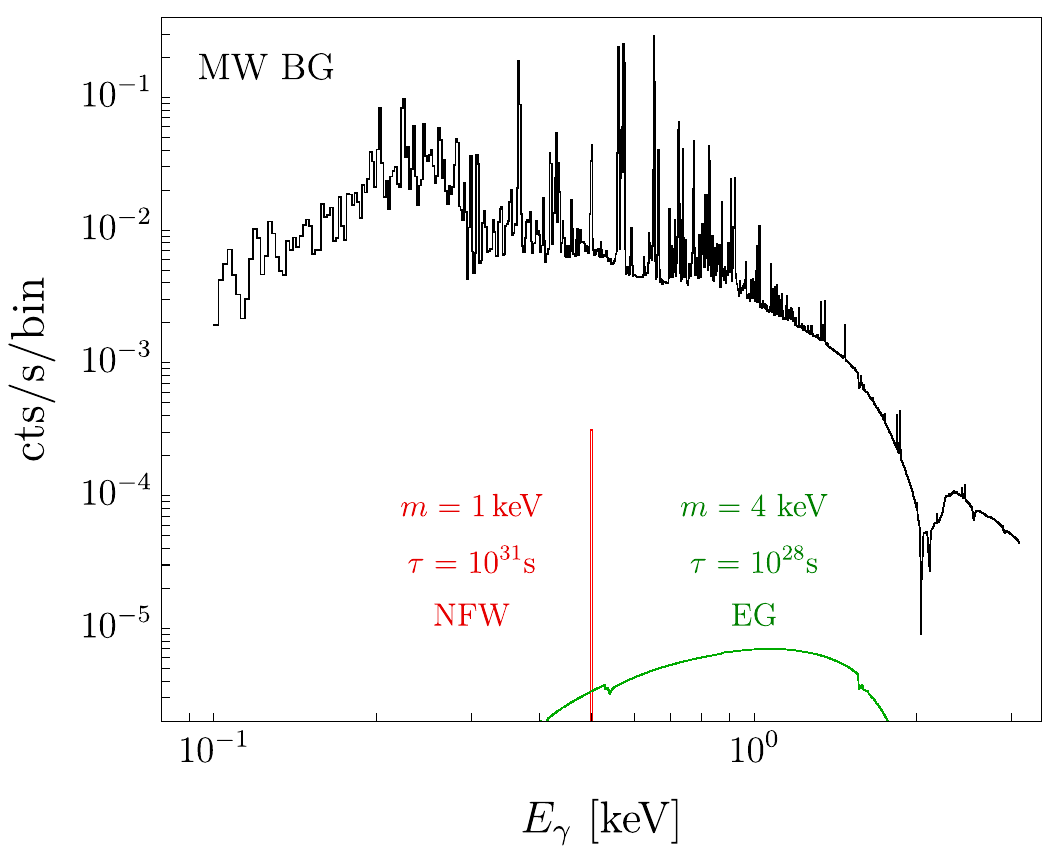}~~~
     \includegraphics[width= 0.442\linewidth]{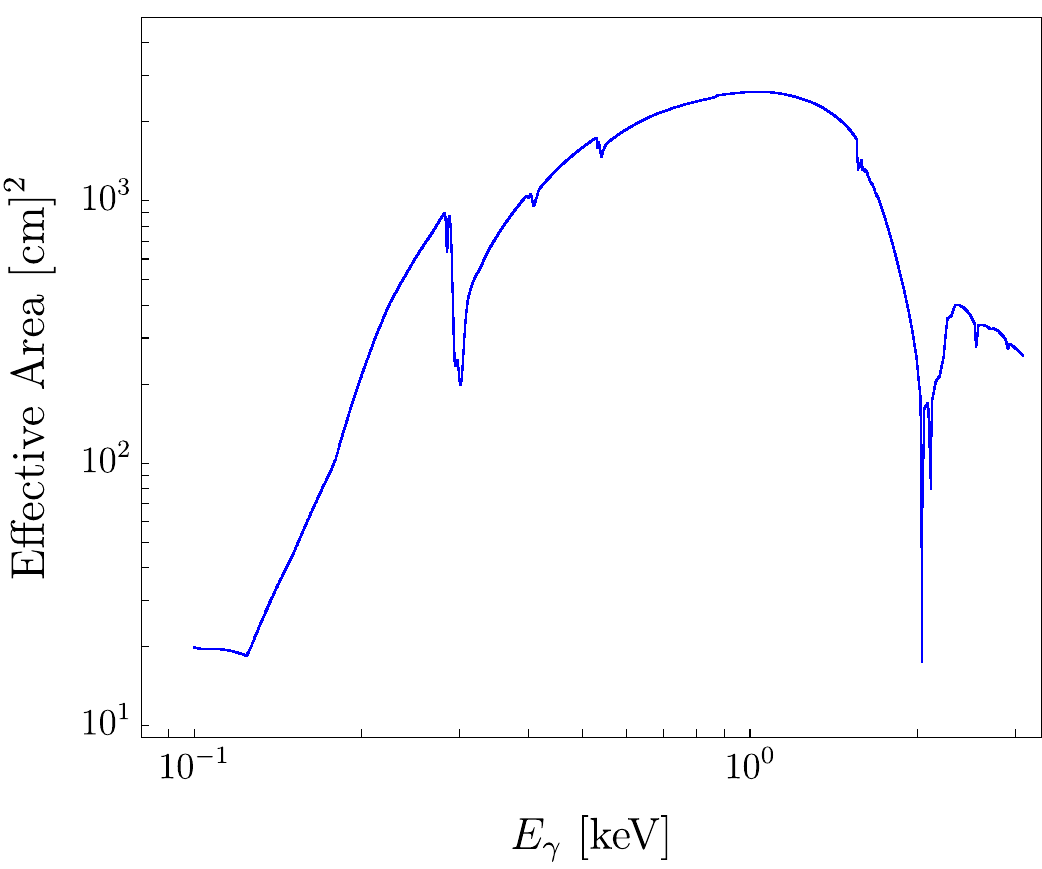}
    \caption{
    {\bf Left:} Background of X-ray emission, which includes 
    both Milky Way emission and contributions from the Cosmic X-ray Background, which is mainly due to extragalactic active galactic nuclei. This emission is 
    convolved with the
    effective area with 2 eV bins to generate the curve shown here (black histograms).  Also shown are two narrow signal curves corresponding
    to representative benchmark points for Galactic decays with an NFW profile (red) and extragalactic (green) emission near the exclusion region shown in Fig. \ref{fig:lifetime}.
  {\bf Right:} Projected effective area of the LEM telescope used in our 
  analysis taken from Ref. \cite{Kraft:2022mnh}.
    }
    \label{fig:foreground}
\end{figure*}

The flux contribution from extragalactic DM decays integrated over all redshifts can be written \cite{Essig:2013goa} 
\be
\label{eq:eg_obs}
\brac{d\phi}{dE_\gamma}_{\rm eg} = \frac{ \Omega_{\rm dm } \, \rho_c}{ 2\pi H_0  m^2 \tau   }   \! \sqrt{\frac{2E_\gamma}{m}} \frac{\Delta \Omega   }{\sqrt{ 
\Omega_m
\! + \! \Omega_\Lambda (2E_\gamma/m)^3}},~~
\ee
where $H_0 = 67.66$ km s$^{-1}$ Mpc$^{-1}$ is the present day Hubble constant, $\rho_c = 4.83 \times 10^{-6} \,\text{GeV cm}^{-3}$  is the critical density, $\Omega_i \equiv \rho_i/\rho_c$,
$\Omega_{\rm dm} = 0.2587$ is the fractional DM density, $\Omega_{\rm m} = 0.3111$ is the fractional matter density, and $ \Omega_\Lambda = 1 - \Omega_m \approx 0.68$ is
the dark energy density \cite{Planck:2018vyg}.

\section{Background Model}
\label{sec:bg-model}

In order to calculate the LEM sensitivity to DM X-ray decays, we need
to estimate the background emission from Galactic and extragalactic sources in the instrumental field of view.  In particular, the soft X-ray background is dominated by the Milky Way interstellar medium and by the cosmic X-ray background (CXB) \cite{2002ApJ...576..188M}. The former is dominated by the emission from the hot gas, while the latter is mainly associated to distant active galactic nuclei \cite{2000Natur.404..459M}. The brightest Galactic emission lines in the energy range of interest are the O VII triplet, the Lyman alpha lines of O VIII and the faint Fe XVII \cite{2002ApJ...576..188M, Kraft:2022mnh}. 

Our emission model taken from Ref. \cite{irina} is shown in Fig. \ref{fig:foreground} (left panel) contains X-ray spectra computed from the Local Hot Bubble and Hot Halo  \cite{2002ApJ...576..188M}. The Local Hot Bubble is modeled using an unabsorbed APEC model \cite{2001ApJ...556L..91S} with temperature $T \approx 0.01$ keV. The Hot Halo component is modeled with two absorbed APEC components with $T =$ 0.225 keV and $T=$ 0.7 keV. Note that the existence of the hotter component of the Hot Halo model is suggested by HaloSat observations \cite{2022HEAD...1940002B}. The emission from the CXB follows the description from Ref. \cite{Hickox_2006} which can be approximated with an absorbed power-law with slope 1.47, whose normalisation is calculated assuming half of the CXB sources are removed. 

In addition to the Milky Way and CXB, we also add a constant detector background rate of 1 event s$^{-1}$ keV$^{-1}$  \cite{irina},
which corresponds to an additional $N_{\rm bg, det} = 4 \times 10^4$ counts in each 2 eV bin, assuming a 20 Ms observation time. This contribution
is not shown in Fig. \ref{fig:foreground}, but is included in our background estimates and is subdominant to other backgrounds for 
all energies below $\sim 1$ keV.

\section{Results}
To evaluate the LEM sensitivity to DM decays, we consider the following test statistic:
\begin{equation}
    \chi^2(m,\tau) =  \sum_{i=1}^{N_{\rm{bin}}}\brac{ N^i_{\rm{sig}} }{\sigma_i}^2~,
    \label{chi}
\end{equation}
% \begin{equation}
% \chi^2(m,\tau) =  \sum_{i=1}^{N_{\rm{bin}}} \left(\dfrac{ {\rm max} \left(N^i_{\rm{sig}} - N^i_{\rm{bg}} \, , 0 \right)}{\sigma_i} \right)^2~,
% \label{chi}
% \end{equation}
where the sum runs over the number of energy bins $N_{\rm{bin}}$ and
 the theoretical counts expected from DM in the $i^{\rm th}$ bin are
\be
N^i_{\rm{sig}} = t_{\rm obs}  \int_{E_i}^{E_{i+1}} dE_\gamma A_{\rm eff}(E_\gamma )\frac{d\phi}{dE_\gamma}(E_\gamma )~,
\ee
  where $t_{\rm obs} = 20$ Ms is the LEM observation time, $A_{\rm eff}$ 
 is the corresponding effective area taken from Ref. \cite{Kraft:2022mnh}, the bin size is 2 eV, and we separately
 calculate LEM projections using the flux from \Eq{eq:therm_obs} for 
 the Galactic contribution
 and \Eq{eq:eg_obs} for the extragalactic contribution. 
 % Here the number of ``observed events" $N_{\rm{obs}}$ is either taken to be 0 (conservative) or to the number of background simulated counts (realistic).
 
 The conservative scenario corresponds to assuming that all detected lines arise from DM decays.
 In the realistic scenario we include the presence of emission lines due to the circumgalactic medium and intergalactic medium (the diffuse gas surrounding and between galaxies). The distribution of events observed by a telescope can be modeled with a binomial distribution. The standard deviation in each energy bin $\sigma_i$ for a Poisson distribution is $\sqrt{N^i_{\rm{bg}}}$,
 where $N^i_{\rm bg}$ includes the background count in the $i^\text{th}$ bin
 taken from Fig. \ref{fig:foreground} (scaled up by 20 Ms observing time) and the constant $N_{\rm bg,det}$ detector background mentioned above.
 % the  Eq. \eqref{chi} corresponds to a chi squared which includes only the energy bins where the DM signal is expected to be above the simulated astrophysical background. 
 For each DM mass, this $\chi^2$ has one degree of freedom, so to obtain LEM projections for the decay lifetime, we perform the Pearson's chi squared test and require $\chi^2 = 4$(25) for $2\sigma$ exclusion  ($5 \sigma$ discovery) sensitivity. Our main result is presented in Fig. \ref{fig:lifetime} in
 the $\tau$ vs. $m$ plane assuming two X-rays are produced per DM decay. 

In the left panel of Fig. \ref{fig:models} we interpret our projections from Fig. \ref{fig:lifetime} in the 
context of an axion-like particle $a$ that decays via $a\to \gamma \gamma$. The lifetime in this 
scenario can be written \cite{Cadamuro:2011fd}
\be
\label{eq:tau_alp}
\tau = \frac{  64 \pi }{ g^2_{a\gamma\gamma} m_a^3} \sim 10^{30} {\rm s}  \brac{10^{-17} \rm GeV^{-1} }{g_{a\gamma\gamma}}^2 \brac{\rm keV}{m_a}^3,~~
\ee
where $g_{a\gamma\gamma}$ is the diphoton coupling. Similarly, in the right panel, we interpret
our results in the context of a sterile neutrino $\nu_s$ which can decay into active species via $\nu_s \to \nu \gamma$, yielding
a single X-ray per decay event. The sterile neutrino radiative width to photon lines is \cite{Palazzo:2007gz}
\be
\label{eq:tau_nus}
\Gamma_\gamma  = \!  \frac{9\alpha G_F^2 \sin^2(2\theta) m_{\nu_s}^5}{1024 \pi^4 }  \sim
10^{-32} \, {\rm s^{-1}} \! \brac{\sin^2(2\theta)}{10^{-10} } \!
\brac{m_{\nu_s}}{\rm keV}^5 \!\!,~~ \nonumber
\ee
where $\alpha$ is the fine-structure constant, $G_F$ is the Fermi constant, $\theta$ is the mixing angle and the lifetime satisfies $\tau^{-1} = \Gamma_{\gamma}/{\rm Br(\nu_s \to \nu \gamma)}$. 
Note that for sterile neutrinos below $m_{\nu_s}\lesssim 164$ eV \cite{Boyarsky:2008ju},  
Tremaine-Gunn bound \cite{PhysRevLett.42.407} excludes identical fermions from being all of the DM.
Furthermore, if sterile neutrinos are produced through their interactions with Standard Model particles
in the early universe, there are also model-dependent limits from structure formation that exclude $m_{\nu_s} \lesssim$ few keV (e.g. \cite{DES:2020fxi}), which vary based on cosmological history and sterile neutrino production mechanisms.
 
\section{Summary}

In this {\it Letter}, we have demonstrated that the LEM telescope can greatly improve the sensitivity to DM photon line decays. We find that, due to the instrument's large grasp and high spectral resolution, a suitable survey can probe lifetimes of order $\sim 10^{32}$ s, exceeding current limits by four orders of magnitude in the few 100 eV $-$ few keV mass range with an observation time of order 10 Ms. Since no instrument has ever been sensitive to DM lifetimes above $\sim 10^{30}$ s in any decay channel, LEM is poised to explore the longest DM lifetimes ever probed.
In light of these promising results, 
LEM may also have sensitivity to secondary X-rays from DM decay and annihilation
to charged particles, which we leave for future work.

\bigskip
\bigskip

\begin{acknowledgments}
We thank Irina Zhuravleva for many helpful conversations and for providing us with the LEM background model.
We also thank Carlos Blanco, Marco Cirelli, Dan Hooper, Alex Kusenko, and Ben Safdi for helpful conversations, and we thank Keyer Thyme for collaboration during the early stages of this work. This work is supported by the Fermi Research Alliance, LLC under Contract No.~DE-AC02-07CH11359 with the U.S. Department of Energy, Office of Science, Office of High Energy Physics. This work was performed in part at the Aspen Center for Physics, which is supported by the National Science Foundation grant PHY-2210452.
\end{acknowledgments}

% For when building bibliography
\bibliography{biblio}

\end{document}